\documentclass[
aps,
 ,twocolumn
]{revtex4}
\usepackage{graphicx}

\begin{document}

\title{Quantum Communication with an Accelerated Partner
}

\author{T.G.Downes, T.C.Ralph and N.Walk}\affiliation{
School of Mathematics and Physics, University
of Queensland, Brisbane, Queensland 4072, Australia}

\date{\today}

\begin{abstract}
{An unsolved problem in relativistic quantum information research is how to model efficient, directional quantum communication between localised parties in a fully quantum field theoretical framework. We propose a tractable approach to this problem based on calculating expectation values of localized field observables in the Heisenberg Picture. We illustrate our approach by analysing, and obtaining approximate analytical solutions to, the problem of communicating coherent states between an inertial sender, Alice and an accelerated receiver, Rob. We use these results to determine the efficiency with which continuous variable quantum key distribution could be carried out over such a communication channel.}

\end{abstract}


\maketitle


\vspace{10 mm}

The study of how information can be carried and processed by systems described by relativistic quantum mechanics is referred to as relativistic quantum information \cite{PER04}. This rapidly growing body of work studies how quantum information tasks and resources are altered by the relativistic treatment of space-time.
A key topic is the sharing of quantum information between inertial and non-inertial parties, as in this case the quantum ground states of the two observers will differ \cite{BIR82} and novel phenomena such as Unruh \cite{UNR76} and Hawking radiation \cite{HAW75} can emerge. An obvious place in which such effects could play a role is in quantum communication tasks, such as quantum key distribution \cite{GIS02}.  However, a rigorous, quantum field theoretic description of such protocols has not been developed. Previous analysis has been restricted to toy models that are ultimately inconsistent with the properties of real communication systems.

In this Letter we introduce a rigorous, and tractable framework for studying optical quantum communication between inertial and non-inertial observers. We specifically apply our approach to the problem of continuous variable quantum communication \cite{BAC03}. We obtain approximate analytical results for conditions typical of such communication systems and use them to analyse quantum key distribution between an inertial Alice and a uniformly accelerating Rob. We find the secret key rate is limited in a time dependent manner.

Traditional detection schemes analysed with acceleration have utilized the Unruh-Dewitt detector \cite{UNR76}. The Unruh-Dewitt detector is a single two-level quantum system weakly coupled to the field over $4 \pi$ steradians. It is not a good model for the efficient, uni-directional macroscopic detectors commonly employed in quantum communication experiments. 
The other method used to describe quantum information protocols in the presence of non-inertial motion does so directly in terms of the field modes \cite{ALS03}, \cite{FUE05}. This approach suffers from two major problems: (i) the restriction of the description to a particular set of unphysical modes, the Unruh modes \cite{UNR76}, in order to simplify the problem; 
and (ii) the use of non-local states defined on single frequency global modes and the subsequent unfounded interpretation of these non-local results in terms of local observers. Although some work has been done on avoiding the latter problem \cite{BRU10} the reliance on the Unruh modes remained. The method developed here avoids both of these problems and leads to a richer and more realistic description of the physics. 

We use a $(3 + 1) D$ massless scalar field description to model a localized, directional, inertial source using Minkowski modes and similarly, a localized, directional, uniformly accelerating detector using Rindler modes. Minkowski co-ordinates, $(x_1, x_2, x_3, t)$, are the standard ones for describing inertial observers. Rindler co-ordinates, $(\xi, x_2, x_3, \tau)$, can be used to describe accelerated observers. The two coordinate systems are related within the right hand sector (the right Rindler wedge - see Fig.1) via \cite{BIR82}:
\begin{eqnarray}
t &= & a^{-1} e^{a \xi} sinh(a \tau); \;\;\;\;\;\;\; x_1 = a^{-1} e^{a \xi} cosh(a \tau); 
\label{R1}
\end{eqnarray}
A stationary observer in Rindler co-ordinates, sufficiently well localized around $\xi = 0$, follows a uniformly accelerated trajectory in Minkowski co-ordinates, such as is depicted in Fig.1. The rate of acceleration is given by the parameter $a$.

The quantum source is held by Alice, who is stationary in Minkowski co-ordinates. The quantum detector is held by Rob, who is stationary in Rindler co-ordinates. We quantify the communication channel between Alice and Rob in terms of expectation values of localized observables, calculated in the Heisenberg Picture, rather than 
analysing the quantum states. It is due to the difficulties of transforming the states that Unruh modes 
have been exclusively employed previously. By analysing the observables we are able to avoid this difficulty and transform between arbitrary modes with ease, leading to the key advantage of this method. 

\begin{figure}[htb]
\begin{center}
\includegraphics*[width=8cm]{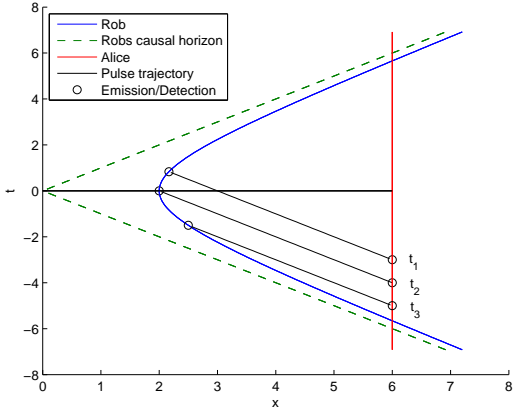}
\caption{Geometry of the quantum communication scenario considered. Alice is stationary, whilst Rob is uniformly accelerating. Alice prepares coherent state pulses and sends them along with local oscillator pulses to Rob at various times (e.g. $t_1, t_2, t_3$). Rob's detector is a broadband, time-integrated homodyne receiver. 
}
\label{fig1}
\end{center}
\end{figure}

To illustrate our approach we consider the following protocol to transmit quantum information from Alice to Rob. Alice sends an optical pulse prepared in a coherent state of amplitude $\alpha$, where $\alpha$ is a complex number, as her signal state. Creation of a coherent state can be modelled as a unitary displacement of the vacuum. Physically, a coherent state is an excellent approximation to the state produced by a well-stabilized laser. She also produces another coherent state of amplitude $\beta$, where $\beta$ is real and $\beta >> |\alpha|$, as a local oscillator mode. She sends both to Rob. Example trajectories of Alice and Rob and the signals sent are depicted on a space-time diagram in Fig.1. Rob performs homodyne detection on the signal and local-oscillator mode, as seen in his reference frame. Homodyne detection enables measurement of the optical quadrature amplitude observables of the signal pulse. The output of the detector at some time $\tau$ (as measured in Rob's frame) can be represented by the following operator \cite{BAC03}:
\begin{equation}
\hat{O}(\tau) = \hat b_i^S \hat b_i^{L\dagger} e^{i \phi} +\hat b_i^{S\dagger}\hat b_i^L e^{-i \phi}
\label{O}
\end{equation}
where $\hat b_i^K$ ($\hat b_i^{K\dagger}$) are boson annihilation (creation) field operators, with $K = S, L$. The superscripts $S,L$ refers to the signal and local-oscillator components of the detected modes. The relative phase $\phi$ determines the quadrature angle detected. The operators can be spectrally decomposed as:
\begin{eqnarray}
\hat{b}_i^K = \int dk_d f_i(k_d) \hat{b}_{k_d}^K \label{rinn}
\end{eqnarray}
For these distributions the $k_d=(k_{d1},k_{d2},k_{d3})$ refers to Rob's detector wave-vector with the first component (corresponding to the direction of acceleration) being the Rindler frequency and the other two components being Minkowski. The integral $\int dk_d$ is over the whole wave-vector space. In this case it is $\int_0^{\infty}\int_{-\infty}^{\infty}\int_{-\infty}^{\infty}dk_{d1}dk_{d2}dk_{d3}$ as the first component of this wave-vector is a right Rindler mode which is strictly positive \cite{TAK86}. The operators $\hat{b}_{k_d}$ are the plane wave Rindler operators. They describe plane waves as perceived by a collection of non-localised accelerating observers. The function $f_i(k_d)$ is what localises these modes in some region of spacetime and hence the modes $\hat{b}_i$ becomes modes perceived by a local accelerating observer. The indices then describe parameters corresponding to these localised observers, for example the central spacial and frequency maxima of the modes and the observation time $\tau$. Rob will integrate the photocurrent from his detector over a time long compared to the pulse length of Alice's signal, thus the average value of the signal received by Rob will be given by the expectation value:
\begin{equation}
X = \langle \int d \tau \hat{O}(\tau) \rangle
\label{X}
\end{equation}
We will also be interested in the variance of the integrated signal
\begin{equation}
V = \langle (\int d \tau \hat{O}(\tau))^2 \rangle - \langle \int d \tau \hat{O}(\tau) \rangle^2
\label{V}
\end{equation}

In Minkowski co-ordinates, the state Alice produces can be represented by displacement of the Minkowski vacuum state, $|0\rangle_M$, as
\begin{equation}
|\alpha,\beta,t\rangle_j=D_j^S(\alpha)D_j^L(\beta)|0\rangle_M 
\label{state}
\end{equation}
where the displacement operators are given by: $D_j^{K}(\gamma)=\exp[\gamma a_j^{K \dagger}-\gamma^*a^{K}_j] $
%
%
with $\gamma=\alpha, \beta$ and 
\begin{eqnarray}
\hat a_j^K&=&\int dk_s f_j(k_s)\hat a_{k_s}^K 
\label{sig}
\end{eqnarray}
Alice's source wave-vector is $k_s=(k_{s1},k_{s2},k_{s3})$. 

In order to calculate the expectation values of Eq.\ref{X} and \ref{V} we need to rewrite Bob's measurement operators in terms of Minkowski modes. The transformation relations between Rindler and Minkowski modes is given by \cite{TAK86}
\begin{eqnarray}
\hat{b}_i^K &=& \int dk_d \int dk_s f_i(k_d) ( A_{k_dk_s} \hat{a}_{k_s}^K + B_{k_dk_s} \hat{a}_{k_s}^{K \dagger}) 
\label{rob}
\end{eqnarray}
where
\begin{eqnarray}
A_{k_dk_s} &=& \frac{\delta(\vec{k}_d-\vec{k}_s)}{\sqrt{2\pi\omega_s(1-e^{-2\pi k_{d1}})}}\left(\frac{\omega_s+k_{s1}}{\omega_s-k_{s1}}\right)^{i\frac{1}{2}k_{d1}} \nonumber \\
B_{k_dk_s} &=& \frac{\delta(\vec{k}_d+\vec{k}_s)}{\sqrt{2\pi\omega_s(e^{2\pi k_{d1}}-1)}}\left(\frac{\omega_s+k_{s1}}{\omega_s-k_{s1}}\right)^{i\frac{1}{2}k_{d1}} \label{bcb}
\end{eqnarray}
are the Bogolyubov coefficients between single frequency Rindler and Minkowski modes where $\vec{k}_d=(k_{d2},k_{d3}), \vec{k}_s=(k_{s2},k_{s3})$ and $\omega_s$ is the source frequency $\omega_s=\sqrt{k_{s1}^2+k_{s2}^2+k_{s3}^2}$. It is straightforward to show that
\begin{eqnarray}
&&D_j^{K \dagger}(\gamma) \hat{b}_i^K D_j^K(\gamma)= \hat{b}_i^K +  \gamma \int dk_d \int dk_s f_i(k_d) \times \nonumber \\
&&\;\;\; \;\;\;\;\;\;\;\;\;\;\;\;\;\;\;\;\;\;\;( A_{k_dk_s} f_j(k_s)^* + B_{k_dk_s} f_j(k_s)) 
\label{exp}
\end{eqnarray}
Substituting Eqs \ref{rob} and their Hermitian conjugates into Eq.\ref{O} and using Eq.\ref{state} as the initial state the expressions for $X$ and $V$ can be expanded via Eq.\ref{exp}. The resulting expressions comprise expectation values of Heisenberg Picture operators over the initial Minkowski vacuum state. Hence we can obtain exact formal solutions for the average quadrature values and their variances. For example the expression for $X$ becomes
\begin{eqnarray}
&& X = \beta(\alpha e^{i \phi} +\alpha^* e^{-i \phi})  \int d\tau \times \;\;\;\;\;\;\;\;\;\;\;\;\;\;\;\;\;\;\;\;\;\;\;\;\;\;\;\;\;\;\;\;\;\;\;\;\; \nonumber \\
&&| \int dk_d \int dk_s f_i(k_d) ( A_{k_dk_s} f_j(k_s)^* + B_{k_dk_s} f_j(k_s))|^2 
\label{exp2}
\end{eqnarray}
where we have used that $\langle 0|_M \; \hat{b}_i^K \;  |0 \rangle_M = 0$. Expressions such as Eq.\ref{exp2} can be numerically solved for specific localized detection and signal wave functions, $f_i(k_d)$ and $f_j(k_s)$ respectively. To obtain analytical solutions we need to make some approximations based on the form of the wave functions.

We assume that the communication between Alice and Rob is ``beam-like" in the sense that Alice sends a well directed Gaussian mode to Rob whom focuses it down to perfectly match the transverse spatial profile of his detector. Moreover, for simplicity, we assume the communication is 
alligned with the acceleration. Hence we can make the paraxial approximation of factoring the signal wave function into transverse and longitudinal components, i.e. $f_j(k_s) = e^{-i (\omega_s t - k_{s1}x)} f_j(k_{s1}) f_j(k_{s2},k_{s3})$ where the origin of Alice's pulse is centred on the space-time point $(x, x_2, x_3, t)$. We later make the assumption that the transverse components of the source match those of the detector wave function (which is similarly factored). Initially we make the impractical assumption that this is achieved with unit efficiency but relax this in our final discussion. 

We also assume that the longitudinal part of the signal wave function is peaked at a large wave number $k_{so}$ such that, for the region of wave numbers for which the wave function is non-zero, $|k_{s1}| >> |k_{s2}|,|k_{s3}|$. We also assume that the standard deviation of the longitudinal part of the wave function, though broad on the wavelength scale is small compared to $k_{so}$. Hence we write $k_{s1} = k_{so} + \bar k$ where $|k_{so}| >> |\bar k|$ for the region of wave numbers for which the wave function is non-zero. These are typical approximations used for non-relativistic quantum communication systems. Given this the longitudinal part of the signal wave function becomes $e^{i |k_{s1}|(\pm x-t)} f_j(k_{s1})$, where $+$ ($-$) corresponds to positive (negative) $k_{so}$. Similarly, we can approximate the signal frequency dependent term in Eqs \ref{bcb} as
\begin{eqnarray}
&& \left(\frac{\omega_s+k_{s1}}{\omega_s-k_{s1}}\right)^{i\frac{1}{2}k_{d1}} \approx 
e^{\pm i\frac{1}{2}k_{d1}(2 ln (2 |k_{s1}|)- ln (k_{s2}^2+k_{s3}^2))} \nonumber \\
&& \;\;\;\;\;\;\;\;\;\; \approx e^{\pm i|\frac{k_{s1}}{k_{so}}|k_{d1}} e^{\pm i k_{d1}(ln (2 |k_{so}|)-\frac{1}{2} ln (k_{s2}^2+k_{s3}^2) -1)}. \nonumber \\
\;\;
\label{approx}
\end{eqnarray}
As a specific case we take $k_{so}<0$ in the following as per the example of Fig.1.

We now turn to the detector wave function. We also factor this into transverse and longitudinal components as: $f_i(k_d) = e^{-i k_{d1}a \tau} f_i(k_{d1}) f_i(k_{d2},k_{d3})$ where the detector is centred on the space-time point $(\xi=0, x_2, x_3, \tau)$. It is important that the longitudinal component of the detector wave function is well localized otherwise its interpretation as a detector following a particular space-time trajectory is compromised. Thus we consider a detector wave function that is very broad in $k_{d1}$ such that it is well localized spatio-temporally. In particular we take $f_i(k_{d1}) \approx 1/\sqrt{2 \pi} $ for $k_{d1} > 0$ and zero otherwise. Substituting our approximate forms into Eq.\ref{exp2} we obtain
\begin{eqnarray}
&&X = \beta(\alpha e^{i \phi} +\alpha^* e^{-i \phi}) \int d\tau \nonumber \\
&& | \int dk_d \int dk_{s1} {{e^{-i k_{d1}a \tau} }\over{\sqrt{2 \pi}}} {{f_i(k_{d2},k_{d3})}\over{\sqrt{2 \pi |k_{so}|(1-e^{-2 \pi k_{d1}})}}} \nonumber \\
&& ( f_j(k_{s1})^* f_j(k_{d2},k_{d3})^* e^{-i|\frac{k_{s1}}{k_{so}}|k_{d1}} \nonumber \\
&& e^{-i k_{d1}(ln 2 |k_{so}|-\frac{1}{2} ln (k_{d2}^2+k_{d3}^2) -1)}e^{i |k_{s1}|(x+t)} + \nonumber \\
&& f_j(k_{s1}) f_i(-k_{d2},-k_{d3}) e^{-i|\frac{k_{s1}}{k_{so}}|k_{d1}}  \nonumber \\
&& e^{-i k_{d1}(ln 2 |k_{so}|-\frac{1}{2} ln (k_{d2}^2+k_{d3}^2) -1)}  e^{-2 \pi k_{d1}} e^{-i |k_{s1}|(x+t)})|^2 \nonumber \\
\;\;
\label{exp3}
\end{eqnarray}
A major simplification of this expression is possible if we assume that $f_j(k_{s1})$ is sufficiently broad in frequency that
\begin{eqnarray}
\int dk_{s1} {{1 }\over{\sqrt{2 \pi |k_{so}|}}} f_j(k_{s1}) e^{-i|\frac{k_{s1}}{k_{so}}|k_{d1}} e^{\pm i |k_{s1}|(x+t)} &\approx& \nonumber \\
\bar f_j  \delta(k_{d1} \mp |k_{so}|(x+t)) 
\label{delta}
\end{eqnarray}
where $\bar f_j$ is the average value of $f_j(k_{s1})$. Firstly, the second term in the absolute square in Eq.\ref{exp3} goes to zero as $(x+t) > 0$ for the range of Alice's source positions considered (see Fig.1) but $f_i(k_{d1}) = 0$ for $k_{d1} < 0$ in the right Rindler wedge. Next, the integral over $\tau$ produces a delta function between the two integrals over $k_{d1}$ (from the absolute square). Inserting the assumption that the transverse components of the source and detector wavefunctions are matched such that $f_j(k_{d2},k_{d3}) = f_i(k_{d2},k_{d3})  e^{i |k_{so}|(x+t)\frac{1}{2} ln (k_{d2}^2+k_{d3}^2)}$, and using the
normalization of the transverse wave function $\int dk_{d2} \int dk_{d3} |f_i(k_{d2},k_{d3})|^2 = 1$, Eq.\ref{exp3} reduces to
\begin{eqnarray}
X &=& \beta(\alpha e^{i \phi} +\alpha^* e^{-i \phi})   \int dk_{s1}'{{ f_j(k_{s1}') \bar f_j^* }\over{(1-e^{-2 \pi |k_{so}|(x+t)})}} \nonumber \\
& \approx & {{ \beta(\alpha e^{i \phi} +\alpha^* e^{-i \phi}) }\over{(1-e^{-2 \pi |k_{so}|(x+t)})}} 
\label{exp4}
\end{eqnarray}
where we have also used an approximate normalization over the signal wave function. By dividing out the amplitude of the local oscillator, $\bar \beta =\beta \sqrt{(1-e^{-2 \pi |k_{so}|(x+t)})}$, we obtain the expectation value of the quadrature amplitude of Alice's signal as observed by Rob
\begin{eqnarray}
\langle X_B(\theta) \rangle_A = X/ \bar \beta
& =& {{ \alpha e^{i \phi} +\alpha^* e^{-i \phi} }\over{\sqrt{(1-e^{-2 \pi |k_{so}|(x+t)})}}}  
\label{exp5}
\end{eqnarray}
Using a similar sequence of approximations we obtain the variance of the signal quadrature as
\begin{eqnarray}
\langle \Delta X_B(\theta)^2 \rangle_A = V/ \bar \beta^2
& =& {{(1+e^{-2 \pi |k_{so}|(x+t)})}\over{(1-e^{-2 \pi |k_{so}|(x+t)})}} 
\label{exp6}
\end{eqnarray}
Eqs \ref{exp5} and \ref{exp6} are our main results, characterising the quadrature signals observed by Rob in the ideal limit of unit efficiency, well localized detection of coherent states sent by Alice. 
Generalization to the detection of other initial states - squeezed states, entangled states, etc is straightforward. 

The solutions have the general form of linear amplification of the initial state, 
as anticipated from single mode treatments \cite{ADE07}. However the effective gain, $G = 1/(1-e^{-2 \pi |k_{so}|(x+t)})$, exhibits a novel dependence on the time at which the pulse is sent, and hence on Rob's position on his trajectory at which he receives the pulse. 
The behaviour of the effective gain can be explained in the following way. Consider first the pulse path labeled $t_2$ in Fig.1. This pulse is received by Rob at $t_R=0$ in Minkowski coordinates. According to Eq.\ref{R1}, $t=0$ corresponds to $\tau=0$ in Rindler coordinates and hence (with $\xi=0$) $x_R=1/a$. Given that $x+t$ is a constant for the path (i.e. the path is a geodesic) we have $x+t=x_R +t_R=1/a$. So the effective gain is $G = 1/(1-e^{-2 \pi |k_{so}|/a})$. Linear amplification of the vacuum with this gain gives a thermal state distribution as a function of the detection frequency, $k_{so}$ and the acceleration on the detector trajectory, $a$, as expected from the Unruh effect \cite{UNR76}. 

Now consider signals sent at $t_1$ ($t_3$), see Fig.1. The $t=0$ intercepts for these signals are $1/a' < 1/a$ ($1/a'' > 1/a$). Therefore the effective gain is lower (higher) for these signals. At first this seems surprising as thermalization due to the Unruh effect is predicted to be constant along the detector trajectory. The explanation is that the detection frequency of the homodyne detector is determined by Alice's local oscillator, as observed by Rob. Rob is instantaneously stationary at $t=0$ and so observes the local oscillator at $k_{so}$. However, Rob receives the signals sent at $t_1$ ($t_3$) when moving away from (towards) Alice. Because of Rob's motion, the effective detection frequency is Doppler-shifted to higher (lower) frequencies resulting in the lower (higher) effectives gains. It is straightforward to show in general that for signals that intercept Rob's trajectory at a point where his instantaneous velocity is $v$, $x+t=1/a\sqrt{(1+v)/(1-v)}$ as expected from the Doppler shift.

A quantum communication protocol that can be implemented via the exchange of coherent states in the way described is continuous variable quantum key distribution \cite{GRO02}. Using established techniques \cite{WEE11} and the channel characterization given by Eqs. \ref{exp5} and \ref{exp6} it is straightforward to calculate the secret key rates if Alice and Rob were to implement this protocol. The results of such a calculation are shown in Fig.2. The figure shows that even with the unrealistic assumption of unit efficiency, secret key rates are reduced by the Unruh effect. The reduction is most pronounced at earlier times, when the signals are Doppler shifted  to lower frequencies that are more effected by the thermalization. If Rob's receiver is assumed to have non-unit efficiency by introducing loss in the standard way, then quantum key distribution becomes impossible at sufficiently early times.

The techniques we have introduced allow the rigorous evaluation of relativistic quantum communication protocols in terms of the localized detectors and sources typically used for quantum communication. The results we have derived here directly apply to continuous variable protocols between an inertial and non-inertial observer. They could straightforwardly be generalized to discrete variable protocols by considering localized number state detection by Rob, extending the description from scalar to vector fields, and considering number state creation by Alice. These techniques could also be adapted to treat quantum communication in curved space \cite{BIR82}, or situations in which inertial detectors couple to Rindler modes due to rapid changes in their energy levels \cite{OLS11}. The latter case is of near term experimental interest.
\begin{figure}[htb]
\begin{center}
\includegraphics*[width=8cm]{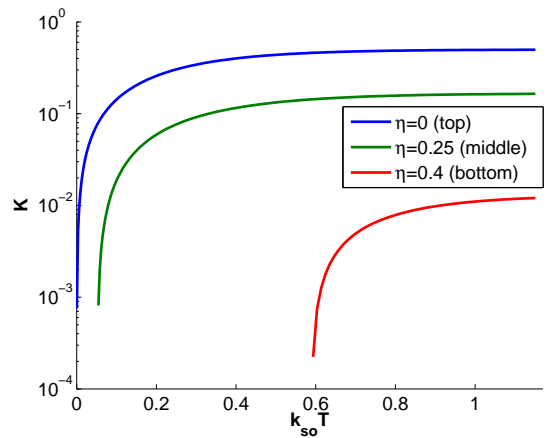}
\caption{Secret key rates obtained for a continuous variable quantum key distribution protocol implemented between Alice and Rob. The key rates ($K$) are plotted as a function of a dimensionless quantity proportional to the emission time, $T =x+t$, and the centre frequency of the pulse, $k_{so}$. Key rates are reduced by a thermal background due to the Unruh effect. Communication efficiency is $1-\eta$. 
}
\label{fig1}
\end{center}
\end{figure}

Acknowledgements: We thank Jorma Louko, Andrzej Dragan and Ivette Fuentes for useful discussions.


\begin{thebibliography}{99}

\bibitem{PER04}
A.~Peres and D.~R.~Terno,
Rev. Mod. Phys. {\bf 76} 93 (2004). 

\bibitem{BIR82} N.D.Birrell and P.C.W.Davies, {\it Quantum fields in curved space}, (Cambridge University Press 1982).

\bibitem{UNR76}
W.~G. Unruh,
Phys. Rev. D {\bf 14} 870 (1976). 

\bibitem{HAW75} S. Hawking, Comm Math Phys {\bf43}, 199 (1975).

\bibitem{GIS02} N.Gisin, et al, 
Rev. Mod. Phys. {\bf 74}, 145 (2002)

\bibitem{BAC03} H-A.Bachor and T.C.Ralph, {\it A guide to experiments in quantum optics}, (2nd Ed, Wiley-VCH, Weinheim 2004).

\bibitem{ALS03}
P.M.Alsing and G.J.Milburn,
Phys Rev Lett {\bf 91}, 180404 (2003). 

\bibitem{FUE05}
I.Fuentes-Schuller and R.B.Mann,
Phys. Rev. Lett. {\bf 95} (2005). 

\bibitem{BRU10} 
D.E.Bruschi, et al, 
Phys. Rev. A {\bf 82}, 042332 (2010).

\bibitem{TAK86} S. Takagi, 
Prog Theor Phys Supp {\bf 88}, 1 (1986); L.C.B.Crispino, et al 
Rev. Mod. Phys. {\bf 80}, 787 (2008).

\bibitem{ADE07} 
G.Adesso, Ivette Fuentes-Schuller, and Marie Ericsson
Phys. Rev. A {\bf 76}, 062112 (2007).

\bibitem{GRO02} 
F.Grosshans, P.Grangier,
Phys.Rev.Lett. {\bf 88}, 057902 (2002).

\bibitem{WEE11} 
C.Weedbrook, et al, 
Rev. Mod. Phys. {\bf 84}, 621(2012).


\bibitem{OLS11} S.Jay Olson and T.C.Ralph, Phys Rev Lett {\bf 106}, 110404 (2011).





\end{thebibliography}
\end{document}